\documentclass[11pt]{article}
\usepackage{color}
\usepackage{subfigure}
\usepackage{amsmath}
\usepackage{amssymb}
\usepackage{graphicx,color}
\usepackage{cite}
\usepackage{enumerate}
\usepackage{amsthm}
\usepackage{amsfonts,mathrsfs}
\usepackage{geometry}
\usepackage{ifthen}
\usepackage{graphicx}
\usepackage{epstopdf}
\usepackage{float}
\usepackage{bm}
\usepackage[caption=false]{subfig}
\usepackage[utf8]{inputenc}

\begin{document}

\title{\bf Coherence dynamics in quantum algorithm for linear systems of equations}

\vskip0.1in
\author{\small Linlin Ye$^1$, Zhaoqi Wu$^1$\thanks{Corresponding author. E-mail: wuzhaoqi\_conquer@163.com},
Shao-Ming Fei$^{2,3}$\\
{\small\it  1. Department of Mathematics, Nanchang University,
Nanchang 330031, P R China}\\
{\small\it  2. School of Mathematical Sciences, Capital Normal University, Beijing 100048, P R China}\\
{\small\it  3. Max-Planck-Institute for Mathematics in the Sciences,
04103 Leipzig, Germany} }

\date{}
\maketitle

\noindent {\bf Abstract} {\small }\\
Quantum coherence is a fundamental issue in quantum mechanics and
quantum information processing. We explore the coherence dynamics of
the evolved states in HHL quantum algorithm for solving the linear
system of equation $A\overrightarrow{x}=\overrightarrow{b}$. By
using the Tsallis relative $\alpha$ entropy of coherence and the
$l_{1,p}$ norm of coherence, we show that the operator coherence of
the phase estimation $P$ relies on the coefficients $\beta_{i}$
obtained by decomposing $|b\rangle$ in the eigenbasis of $A$. We
prove that the operator coherence of the inverse phase estimation
$\widetilde{P}$ relies on the coefficients $\beta_{i}$, eigenvalues
of $A$ and the success probability $P_{s}$, and it decreases with
the increase of the probability when $\alpha\in(1,2]$. Moreover, the
variations of coherence deplete with the increase of the success
probability and rely on the eigenvalues of $A$ as well as the
success probability.

\vskip0.2in

\noindent {\bf 1 Introduction}\\\hspace*{\fill}\\
As one of the most essential features of quantum
physics and the important resources in quantum information
processing, quantum coherence first arose in the field of optics and
was used for studying the interference ability of light
\cite{SMO,GRJ,SEC}, which has applications in many fields such as
transport theory \cite{RPMM,WBM}, biology systems \cite{PMB,HS,LS}
and nanoscale physics \cite{KOL}. Quantification of coherence is
pivotal both theoretically and practically. A seminal framework for
quantifying coherence as a resource has been proposed in \cite{TB},
while a useful and tractable equivalent regime has been given in
\cite{YXD}. Furthermore, quantification of coherence in
infinite-dimensional systems has been considered in \cite{JXP,YRZ}.
Recently, the role of coherence has been widespreadly concerned in
studying the thermodynamic properties of small systems at low
temperatures \cite{LM,MLK,VNG,MHJ}. Great efforts have also been
devoted to establishing the relationships between coherence and
other quantum resources \cite{ECM,JMB,ZXY,YYX}, while the research
on coherence from other perspectives, such as coherence dynamics
\cite{SDZ,TRB,JWZ,YPYJ,WZQH}, distribution of coherence \cite{CRM},
average coherence and coherence generating power of quantum channels
\cite{WZZF,WZZL} have also been carried out.

As a generalization of Shannon entropy, Tsallis
entropy was introduced in \cite{TC}, which has aroused great
interest in the field of statistical physics. Tsallis relative
entropy, acting as an important distance measure, is invariant under
unitary transformation and possesses monotonicity under projective
measurements \cite{FSY}. Tsallis relative $\alpha$ entropy, as the
quantum version, has been examined as a measure of the degree of
state purification \cite{AS2}. Tsallis relative $\alpha$ entropy of
coherence, a quantifier of coherence for quantum states, was first
proposed in \cite{RAEQ}, though it does not satisfy the the strong
monotonicity. By remedying the definition in \cite{RAEQ}, Zhao and
Yu \cite{ZHYC} provided a bona-fide coherence measure. On the other
hand, the definition of the $l_{q,p}$ norm \cite{KAL} only relies on the
space structure of matrices.
It has been proved that the $l_{q,p}$ norm of coherence is a
well-defined quantifier with $q=1$ and $p\in[1,2]$ \cite{JYL}, which
has a simple closed form extending and unifying previous norm-induced coherence measures.

Quantum computers take full advantage of quantum state superposition
and entanglement to improve the speed of problem solving
\cite{MA,ZhouNR} via quantum algorithms. In the case of specific
scenarios, Shor's algorithm \cite{PWSP} and simulations of quantum
systems \cite{LSU,ADS} exhibit exponential acceleration compared to
classical algorithms. Grover's search quantum algorithm can speed up
the search process \cite{GLK}, while Deutsch-Jozsa quantum algorithm
simplifies the function evaluation, and provides a simpler analysis
of the qubits between the use of entanglement in it \cite{CDKK}. In
evaluating the normalized trace of unitary matrices, the
deterministic quantum computation with one qubit (DQC1) model
enables potential exponential quantum speedup \cite{KELR}.
Bernstein-Vazirani algorithm \cite{DJSM} is a two-qubit algorithm in
a single query on an ensemble quantum computer, which allows one to
determine a bit string encoded into an oracle.

It is an important problem to solve linear systems of equations in
science and engineering. To this end, Harrow, Hassidim and Lloyd
\cite{HAW} proposed HHL quantum algorithm with exponential speedup
compared with classical algorithms that solves the problem
$A\overrightarrow{x}=\overrightarrow{b}$, where $A$ is a Hermitian
$s$-sparse $N\times N$ matrix, and $\overrightarrow{b}$ is a unit
vector. However, usually one is not interested in
$\overrightarrow{x}$ itself, but the approximation of the
expectation value of some quantum mechanical operators.

HHL quantum algorithm is a breakthrough in giving efficient
solutions of high-dimensional linear systems. It is different from
some deterministic ones \cite{DDQ} that provide the solution by a
determinate final pure state. For implementing the algorithm, an
efficient and generic quantum circuit design \cite{CYD} for solving
linear systems has been given. Experimental realization \cite{WJK}
of HHL quantum algorithm has been demonstrated in superconducting
system \cite{ZYS}, optical system \cite{PJC} and nuclear magnetic
system \cite{CXD}. In \cite{CBD}, the authors have introduced the
preconditioning of the algorithm and extended its applicability.
Wossnig, Zhao and Prakash \cite{WLZ} have solved the problem that
when HHL quantum algorithm is applied to dense matrices the
potential exponential advantage of it is lost. The dependence of the
algorithm on the condition number has been improved in subsequent
works \cite{SYS,ADL}. This algorithm has been used to implement the
finite element method for $d$-dimensional boundary value problems
\cite{MAP}, and it has inspired several works \cite{RPM,SMS,WNB} in
the emerging research area of quantum machine learning.

It is an interesting problem to explore the role of entanglement,
correlation and coherence played in quantum algorithms as physical
resources, which has attracted much attention in recent years.
Entanglement dynamics and coherence dynamics in Grover's algorithm
and Deutsch-Jozsa algorithm have been investigated extensively
\cite{MinghuaPan2017QIP,MinghuaPan2019TCS,SHL,HMC,LYC}. The
performance of Bernstein-Vazirani algorithm is directly related to
the number of coherence in the initial state in the absence of
entanglement\cite{NMK}, and in Shor's algorithm, coherence is a
quantum resource that gives lower and upper bounds of the
performance \cite{AFTTE}. It is found that the overall effect in
Grover's search algorithm is that the $l_{1}$ norm of coherence is
depleted during the quantum search process \cite{PMQ}, and the
complementarity relation between the success probability and
coherence has been derived \cite{MPH}. Following the idea in
\cite{PMQ,MPH}, coherence dynamics in Grover's search algorithm has
been studied based on Tsallis relative $\alpha$ entropy \cite{YLW}.
Recently, Feng, Chen and Zhao have studied coherence and
entanglement in Grover and HHL algorithm \cite{FCC}. It has been
observed that the DQC1 is directly related to the recoverable
coherence \cite{JMM}, and the uncertainties, coherence and quantum
correlations in DQC1 have also been explored in detail
\cite{FSH,WWH,GEI}. In this paper, we study coherence dynamics in
HHL algorithm based on two important coherence measures, i.e., the
Tsallis relative $\alpha$ entropy of coherence and the $l_{1,p}$
norm of coherence for $p\in[1,2]$.

The remainder of this paper is structured as follows. In Section 2,
we recall HHL quantum algorithm and the coherence quantifiers based
on Tsallis relative $\alpha$ entropy and $l_{1,p}$ norm. In Section
3, we investigate the coherence dynamics of the state after each
basic operator is applied, and formulate the coherence production
and depletion in HHL quantum algorithm. Moreover, we establish the
relationships between the coefficients $\beta_{i}$ and the elements
of the vector $\overrightarrow{b}$. In Section 4, we choose some
specific linear systems of equations to illustrate the coherence
dynamics of the state after each basic operator is applied, and
present the variations of the coherence during the HHL quantum
algorithm application. Some concluding remarks are given in Section
5.

\vskip0.1in

\noindent {\bf 2 HHL quantum algorithm and coherence quantifiers based on Tsallis relative $\alpha$ entropy and $l_{1,p}$ norm }\\\hspace*{\fill}\\
In this section, we recall HHL quantum algorithm and coherence
quantifiers based on Tsallis relative $\alpha$ entropy and $l_{1,p}$
norm. \vskip0.1in

\noindent {\bf 2.1 HHL quantum algorithm}\\\hspace*{\fill}\\
We describe here the basic idea of HHL quantum algorithm. Given an
$s$-sparse $N\times N$ Hermitian matrix with a spectral
decomposition $A=\sum_{j=1}^{N}\lambda_{j}|u_{j}\rangle\langle
u_{j}|$ and a unit vector $\overrightarrow{b}$, where $s$-sparse
indicates that $A$ has at most $s$ nonzero elements per row,
$\lambda_{j}$ is the eigenvalue of $A$ and $|u_{j}\rangle$ is the
corresponding eigenstate. Solving the linear system of equations
$A\overrightarrow{x}=\overrightarrow{b}$ is equivalent to finding
state $|x\rangle$ that satisfies $A|x\rangle=|b\rangle$, where
$|b\rangle$ is the representation of vector $\overrightarrow{b}$ in
the eigenbasis of $A$. For the linear system of equations, the
quantum mechanical solution is $|x\rangle\varpropto
A^{-1}|b\rangle\varpropto\sum_{j=1}\frac{\beta_{j}}{\lambda_{j}}
|u_{j}\rangle$. The main steps of HHL quantum algorithm can be
summarized as follows: transform a given $s$-sparse $N\times N$
Hermitian matrix $A$ into a unitary operator
$\mathrm{e}^{\mathrm{i}At}$, and simulate
$\mathrm{e}^{\mathrm{i}At}$ in time \cite{BDW}
$\widetilde{O}(\log(N)s^{2}t)$, where $\widetilde{O}$ suppresses
items that grow slower.
Represent vector $\overrightarrow{b}$ as a quantum state $|b\rangle=\sum_{i=1}^{N} b_{i} |i\rangle$ and store it in a quantum register termed $B$, where $\sum_{i=1}^{N}|b_{i}|^{2}=1$, with $|i\rangle$ the basis state of the register $B$.\\
$\bullet$ (i) Apply the phase estimation \cite{LAP,BVD}.\\ (1) Initialize the qubits by transforming the first register termed $D$
to the state
\begin{equation}\label{eq1}
 |\psi_{0}\rangle=\sqrt{\frac{2}{T}}\sum_{\tau=0}^{T-1}\sin\frac{\pi(\tau+\frac{1}{2})}{T}|\tau\rangle,
\end{equation}
where $|\tau\rangle$ represents the basis state of the first register $D$. In order to minimize a certain quadratic loss function, we could choose appropriate coefficients \cite{LAP,BVD} of $|\psi_{0}\rangle$.\\
(2) Apply the conditional Hamiltonian evolution
$\sum_{\tau=0}^{T-1}|\tau\rangle\langle\tau|^{D}\otimes
\mathrm{e}^{\mathrm{i}A\tau t_{0}/T}$ on
$|\psi_{0}\rangle^{D}\otimes|b\rangle$, where
$t_{0}=O(\kappa/\epsilon)$, $T=2^{l}$, with $\kappa$ being the
condition number of $A$, or the ratio between the largest and smallest
eigenvalues of $A$, $\epsilon$ is the additive error achieved in
the output state $|x\rangle$, $l$ represents the qubit number of the
first register $D$. \\
(3) Apply the quantum Fourier
transform to the first register $D$. Then we have
\begin{equation}\label{eq2}
\sum_{j=1}^{N}\beta_{j}|\lambda_{j}\rangle|u_{j}\rangle,
\end{equation}
where $\lambda_{j}\approx\frac{2\pi k}{t_{0}}$, $|k\rangle$ is the basis state after the quantum Fourier transform. Via the phase estimation, with the ability to expand $|b\rangle$ in the eigenbasis of $A$ as $|b\rangle=\sum_{j=1}^{N}\beta_{j} |u_{j}\rangle$, one gets the corresponding eigenvalues $\lambda_{j}$ of $A$.\\
$\bullet$ (ii) Add an ancilla qubit and apply the conditional rotation $C$-$R_{y}(\theta_{j})$. Then we have the state
\begin{equation}\label{eq3}
\sum_{j=1}^{N}\beta_{j}|\lambda_{j}\rangle|u_{j}\rangle\left(\sqrt{1-\frac{C^{2}}
{\lambda_{j}^{2}}}|0\rangle+\frac{C}{\lambda_{j}}|1\rangle\right),
\end{equation}
where $C=O(\frac{1}{\kappa})$ is an appropriate constant \cite{CYD}, $R_{y}(\theta_{j})=\mathrm{e}^{-\mathrm{i}\frac{\theta_{j}}{2}Y}=\mathrm{e}
^{-\mathrm{i}Y\arcsin\frac{C}{\lambda_{j}}}$, with $Y$ being the Pauli operator, $\theta_{j}=2\arcsin\frac{C}{\lambda_{j}}$.\\
$\bullet$ (iii) Undo the phase estimation to uncompute the $|\lambda_{j}\rangle$, i.e., apply the inverse phase estimation to restore the register $D$. Then we have
\begin{equation}\label{eq4}
\sum_{j=1}^{N}\beta_{j}|u_{j}\rangle\left(\sqrt{1-\frac{C^{2}}{\lambda_{j}^{2}}}
|0\rangle+\frac{C}{\lambda_{j}}|1\rangle\right).
\end{equation}
$\bullet$ (iv) Measure the last qubit. If it returns 1, then we
obtain
\begin{equation}\label{eq5}
\sqrt{\frac{1}{\sum_{j=1}^{N}C^{2}\beta_{j}^{2}/\lambda_{j}^{2}}}
\sum_{j=1}^{N}\beta_{j}\frac{C}{\lambda_{j}}|u_{j}\rangle.
\end{equation}
It corresponds to
$\sum_{j=1}^{N}\frac{\beta_{j}}{\lambda_{j}}|u_{j}\rangle$ up to
normalization, which is equal to the solution $|x\rangle$  of the
linear system of equations.

The success probability of HHL quantum algorithm is
$P_{s}=\sum_{j=1}^{N}C^{2}\beta_{j}^{2}/\lambda_{j}^{2}$, and since
$C=O(\frac{1}{\kappa})$, the success probability is at least
$\Omega(\frac{1}{\kappa^{2}})$. The prerequisite for using HHL
quantum algorithm to solve the linear system of equations is that
$A$ is a Hermitian matrix. If $A$ is not a Hermitian matrix, we can
define $ Q=\left(\begin{matrix}
0\ A\\
A^{\dagger}\ 0
\end{matrix}
\right).
$
As $Q$ is a Hermitian matrix, we can solve the linear system of
equations
$Q\overrightarrow{y}=\left(\begin{matrix}\overrightarrow{b}\\
0\end{matrix}\right)$ and obtain
$\overrightarrow{y}=\left(\begin{matrix}0\\
\overrightarrow{x}\end{matrix}\right)$. Then we get the solution of
the linear system of equations
$A\overrightarrow{x}=\overrightarrow{b}$ when $A$ is not a Hermitian matrix.

HHL quantum algorithm indicates another promising application of
quantum computers. Since most of other algorithms use a quantum
linear system solver as a subroutine such as data fitting, any
improvement about quantum linear system algorithm is a significant
computational advantage. It is important in determining certain
properties of stochastic processes, and allows us to perform useful
computations such as sampling from the solution vector.

\vskip0.1in

\noindent {\bf 2.2 Tsallis relative $\alpha$ entropy of coherence and $l_{1,p}$ norm of coherence}\\\hspace*{\fill}\\
The Tsallis relative $\alpha$ entropy is a good
information-theoretic measure for pure reference state, which is
defined by \cite{AS1,AS2}
\begin{equation}\label{eq6}
D_{\alpha}(\rho\|\sigma)=\frac{1}{\alpha-1}\left(f_{\alpha}(\rho,\sigma)
-1\right),
\end{equation}
where $\alpha\in(0,1)\cup(1,\infty)$,
$f_{\alpha}(\rho,\sigma)=\mathrm{Tr}(\rho^{\alpha}\sigma^{1-\alpha})$.
It is monotonic under trace-preserving completely positive maps.
Note that $D_{\alpha}(\rho\|\sigma)$ reduces to
$S'(\rho\parallel\sigma)=\ln2\cdot S(\rho\parallel\sigma)$ when
$\alpha\rightarrow1$, where $S(\rho\parallel\sigma)=
\mathrm{Tr}(\rho\log\rho)- \mathrm{Tr}(\rho\log\sigma)$ is the
standard relative entropy that is of fundamental importance to
physics and the logarithm `log' is taken to be base 2. Fixing a set
of orthogonal basis $\{|i\rangle\}_{i=1}^d$ in a $d$ dimensional
Hilbert space, the coherence $\tilde{C}_{\alpha}(\rho)$ based on
Tsallis relative $\alpha$ entropy is defined by \cite{RAEQ}
\begin{equation}\label{eq7}
\tilde{C}_{\alpha}(\rho)=\mathop{\mathrm{min}}\limits_{\sigma\in \mathcal{I}}
D_{\alpha}(\rho\|\sigma)=\frac{1}{\alpha-1}\left[\left(\sum_{i=1}^{d}\langle i|\rho^{\alpha}|i\rangle^{\frac{1}{\alpha}}\right)^{\alpha}-1\right],
\end{equation}
where $\mathcal{I}$ denotes the set of incoherent states that are
diagonal in the given basis. However, $\tilde{C}_{\alpha}(\rho)$
only satisfies monotonicity and variational monotonicity, but does
not satisfy the strong monotonicity. Based on
Tsallis relative $\alpha$ entropy, for $\alpha\in(0,1)\cup(1,2]$, a
well-defined coherence quantifier $C_{\alpha}(\rho)$ \cite{ZHYC} has
been presented,
\begin{equation}\label{eq8}
C_{\alpha}(\rho)=\mathop{\mathrm{min}}\limits_{\sigma\in \mathcal{I}}\frac{1}{\alpha-1}\left(f_{\alpha}^{\frac{1}{\alpha}}(\rho,\sigma)
-1\right)=\frac{1}{\alpha-1}\left[\sum_{i=1}^{d}\langle i|\rho^{\alpha}|i\rangle^{\frac{1}{\alpha}}-1\right].
\end{equation}

Note that when $\alpha\rightarrow1$,
$C_{\alpha}(\rho)$ reduces to $\ln2\cdot C_{r}(\rho)$, where
$C_{r}(\rho)=\mathrm{Tr}(\rho\log\rho)-
\mathrm{Tr}(\rho_{\mathrm{diag}}\log\rho_{\mathrm{diag}})$, with
$\rho_{\mathrm{diag}}$ the diagonal part of the state $\rho$, is the
relative entropy of coherence \cite{TB}. In multipartite systems,
the relative entropy of coherence with free reference basis
constructed the tradeoff relation (monogamy or polygamy), which
depends on the state and accompanied by the basis-free coherence. In
principle, only the relative entropy of coherence can be exactly
measured without the full quantum state tomography for experimental
practice \cite{WAG}.

On the other hand, $C_{\alpha}(\rho)$ reduces to
$2C_{s}(\rho)$ when $\alpha=\frac{1}{2}$, where
$C_{s}(\rho)=1-\sum_{j=1}^{d}\langle j|\sqrt{\rho}|j\rangle^{2}$ is
the skew information of coherence \cite{YCS}, which serve as the
natural upper bound on the quantum discord, and has a close relation
with the corresponding experimental implementations and quantum
metrology.

The $l_{q,p}$ norm of a matrix $A\in M_{n}$ is the $l_{q}$ norm of the vector formed by the $l_{p}$ norm of the columns of $A$,
\begin{equation}\label{eq9}
l_{q,p}(A)=\left(\sum_{j=1}^{n}l_{p}(A_{j})^{q}\right)^{\frac{1}{q}},~~~1\leq p,~q\leq\infty,
\end{equation}
where $A_{j}$ stands for the $j$th column of $A$, and
\begin{equation*}\label{eq}
l_{p}(A_{j})=\left(\sum_{i=1}^{n}|A_{i,j}|^{p}\right)^{\frac{1}{p}},~~~1\leq p,
\end{equation*}
where $A_{i,j}$ is the entry of the $i$th row and $j$th column of
the matrix $A$. Note that $l_{p,p}$ norm is
actually the $l_p$ norm. For simplicity, we use $C_{q,p}$ to denote
$C_{l_{q,p}}$, the coherence induced by $l_{q,p}$ norm, and it is
pointed out that \cite{JYL}
\begin{equation*}
C_{q,p}(\rho)=\mathop{\mathrm{min}}\limits_{\sigma\in\mathcal{I}}l_{q,p}(\rho-\sigma)
\end{equation*}
is a well-defined coherence measure if and only if $q=1$ and
$p\in[1,2]$, which provides a class of potentially useful coherence
measures and extends the techniques to study quantum coherence in
multipartite systems. The coherence $C_{1,p}$ of a density operator
$\rho$ for $p\in[1,2]$ is \cite{JYL}
\begin{align}\label{eq10}
C_{1,p}(\rho)\notag
=&\mathop{\mathrm{min}}\limits_{\sigma\in\mathcal{I}}l_{1,p}(\rho-\sigma)
=l_{1,p}(\rho-\rho_{\mathrm{diag}})=\sum_{j=1}^{n}l_{p}((\rho-\rho_{\mathrm{diag}})_{j})\\
=&\sum_{j=1}^{n}\left(\sum_{i=1}^{n}|(\rho-\rho_{\mathrm{diag}})_{i,j}|^{p}\right)^{\frac{1}{p}},
\end{align}
where $(\rho-\rho_{\mathrm{diag}})_{i,j}$ denotes entry of the $i$th
row and $j$th column of the matrix $\rho-\rho_{\mathrm{diag}}$.
When $p=1$, $C_{1,p}(\rho)$ reduces to the $l_{1}$ norm
of coherence with analytic expression $C_{l_{1}}(\rho)=\sum_{i\neq
j}|\rho_{ij}|$ \cite{TB}, which is simple and direct that induce the
measurable bounds \cite{YCS}. Note that $C_{1,2}$ has better
smoothness properties than the important measure $C_{l_1}$. Note
that some $l_{q,p}$ norms induce coherence measures while no unitary
similarity invariant norm does, since a quantum state can always be
transformed into a diagonal state via a unitary similarity
transformation.

\vskip0.1in

\noindent {\bf 3 Coherence dynamics in HHL quantum algorithm }\\\hspace*{\fill}\\
In this section, we investigate the coherence dynamics of the state
after each basic operator is applied in HHL quantum algorithm, and study
the production and depletion of coherence based
on the Tsallis relative $\alpha$ entropy and the $l_{1,p}$ norm. We explore
the relationships between the coefficients $\beta_{i}$ and the elements
of the vector $\overrightarrow{b}$ in the algorithm.

Denote the phase estimation by $P$, the state after the phase
estimation is given by
\begin{equation*}\label{eq}
|\psi_{P}\rangle=\sum_{j=1}^{N}\beta_{j}|\lambda_{j}\rangle|u_{j}\rangle.
\end{equation*}
Let $\rho_{P}=|\psi_{P}\rangle\langle\psi_{P}|
=\sum_{i,j=1}^{N}\beta_{j}\beta_{i}|\lambda_{j}u_{j}\rangle\langle\lambda_{i}u_{i}|$ be the density operator of the state $|\psi_{P}\rangle$. By employing Eqs. (\ref{eq10}) and (\ref{eq8}) we have\\
$\mathbf{Theorem}$ $\mathbf{1}$ The coherence of the state $|\psi_{P}\rangle$ based on the $l_{1,p}$ norm and the Tsallis relative $\alpha$ entropy are given by
\begin{equation}\label{eq11}
C_{1,p}(\rho_{P})=\sum_{i=1}^{N}\left(\sum_{j\neq i}|\beta_{i}\beta_{j}|^{p}\right)^{\frac{1}{p}}
\end{equation}
and
\begin{equation}\label{eq12}
C_{\alpha}(\rho_{P})=\frac{1}{\alpha-1}\left(\sum_{i=1}^{N}\beta_{i}^{\frac{2}{\alpha}}-1\right),
\end{equation}
respectively.\\\hspace*{\fill}\\
{\bf Remark 1} Setting $p=1$ in Eq. (\ref{eq11}), we obtain the
$l_{1}$ norm of coherence of $\rho_{P}$,
\begin{equation}\label{eq16}
C_{l_{1}}(\rho_{P})=\sum_{i\neq j}|\beta_{i}\beta_{j}|.
\end{equation}
Letting $\alpha=\frac{1}{2}$ and taking the limit $\alpha\rightarrow1$ in Eq. (\ref{eq12}), we have the skew information of coherence and the relative entropy of coherence of $\rho_{P}$,
\begin{equation}\label{eq17}
C_{s}(\rho_{P})=1-\sum_{i=1}^{N}\beta_{i}^{4}
\end{equation}
and
\begin{equation}\label{eq18}
C_{r}(\rho_{P})=-\sum_{i=1}^{N}\beta_{i}^{2}\log\beta_{i}^{2},
\end{equation}
respectively.

According to Theorem 1, the coherence of $\rho_{P}$ based on the $l_{1,p}$ norm and the Tsallis relative $\alpha$ entropy relies on the coefficients $\beta_{i}$ obtained by decomposing $|b\rangle$ in the eigenbasis of $A$. Correspondingly, according to Eqs. (\ref{eq16}), (\ref{eq17}) and (\ref{eq18}), we observe that the $l_{1}$ norm of coherence, the skew information of coherence and the relative entropy of coherence of $\rho_{P}$ all depend on the coefficients $\beta_{i}$.

Denote the conditional rotation $C$-$R_{y}(\theta_{j})$ by $R$. The state after
the conditional rotation $C$-$R_{y}(\theta_{j})$ is given by
\begin{equation*}\label{eq}
|\psi_{R}\rangle=\sum_{j=1}^{N}\sqrt{1-\frac{C^{2}}
{\lambda_{j}^{2}}}\beta_{j}|\lambda_{j}\rangle|u_{j}\rangle
|0\rangle+\frac{C}{\lambda_{j}}\beta_{j}|\lambda_{j}\rangle|u_{j}\rangle|1\rangle.
\end{equation*}
Let $\rho_{R}$ be the density operator of the state $|\psi_{R}\rangle$.
From Eqs. (\ref{eq10}) and (\ref{eq8}) we have by direct calculation\\
$\mathbf{Theorem}$ $\mathbf{2}$ The coherence of the state $|\psi_{R}\rangle$ can be expressed as
\begin{align}\label{eq19}
C_{1,p}(\rho_{R})\notag
=&\sum_{i=1}^{N}\left(\sum_{j\neq i}\left(|\beta_{i}\beta_{j}|\sqrt{1-\frac{C^{2}}{\lambda_{j}^{2}}}
\sqrt{1-\frac{C^{2}}{\lambda_{i}^{2}}}\right)^{p}+\sum_{j=1}^{N}\left(\left|\frac{C\beta_{i}\beta_{j}}
{\lambda_{j}}\right|\sqrt{1-\frac{C^{2}}{\lambda_{i}^{2}}}\right)^{p}\right)^{\frac{1}{p}}\\
+&\sum_{i=1}^{N}\left(\sum_{j\neq i}\left|\frac{C^{2}\beta_{i}\beta_{j}}{\lambda_{i}\lambda_{j}}\right|^{p}+\sum_{j=1}^{N}
\left(\left|\frac{C\beta_{i}\beta_{j}}{\lambda_{i}}\right|
\sqrt{1-\frac{C^{2}}{\lambda_{j}^{2}}}\right)^{p}\right)^{\frac{1}{p}},
\end{align}
based on the $l_{1,p}$ norm, and
\begin{equation}\label{eq20}
C_{\alpha}(\rho_{R})=\frac{1}{\alpha-1}\left[\sum_{i=1}^{N}\beta_{i}^{\frac{2}{\alpha}}
\left(\left(1-\frac{C^{2}}{\lambda_{i}^{2}}\right)^{\frac{1}{\alpha}}+\left(\frac{C^{2}}
{\lambda_{i}^{2}}\right)^{\frac{1}{\alpha}}\right)-1\right],
\end{equation}
based on the Tsallis relative $\alpha$ entropy,
where $C=O(\frac{1}{\kappa})$ is an appropriate constant with $0\leq\frac{C^{2}}{\lambda_{i}^{2}}\leq1$.\\\hspace*{\fill}\\
{\bf Remark 2} Setting $p=1$ in Eq. (\ref{eq19}), we obtain the
$l_{1}$ norm of coherence of $\rho_{R}$ as
\begin{equation}\label{eq23}
C_{l_{1}}(\rho_{R})
=\sum_{i\neq j}\left(|\beta_{i}\beta_{j}|\sqrt{1-\frac{C^{2}}{\lambda_{j}^{2}}}
\sqrt{1-\frac{C^{2}}{\lambda_{i}^{2}}}+C^{2}\left|\frac{\beta_{i}\beta_{j}}
{\lambda_{i}\lambda_{j}}\right|\right)
+2\sum_{i,j=1}^{N}\left|\beta_{i}\beta_{j}\frac{C}{\lambda_{i}}\right|
\sqrt{1-\frac{C^{2}}{\lambda_{j}^{2}}}.
\end{equation}
Letting $\alpha=\frac{1}{2}$ in Eq. (\ref{eq20}), one gets that the skew information of coherence of $\rho_{R}$ can be expressed as
\begin{equation}\label{eq24}
C_{s}(\rho_{R})=1-\sum_{i=1}^{N}\beta_{i}^{4}\left[\left(1-\frac{C^{2}}{\lambda_{i}^{2}}\right)^{2}
+\left(\frac{C^{2}}{\lambda_{i}^{2}}\right)^{2}\right].
\end{equation}
Under the limit $\alpha\rightarrow1$, from Eq. (\ref{eq20}) the relative entropy of coherence of $\rho_{R}$ is given by
\begin{equation}\label{eq25}
C_{r}(\rho_{R})=-\sum_{i=1}^{N}\left[\beta_{i}^{2}\left(1-\frac{C^{2}}{\lambda_{i}^{2}}\right)
\log\beta_{i}^{2}\left(1-\frac{C^{2}}{\lambda_{i}^{2}}\right)+\beta_{i}^{2}\frac{C^{2}}
{\lambda_{i}^{2}}\log\beta_{i}^{2}\frac{C^{2}}{\lambda_{i}^{2}}\right].
\end{equation}

According to Theorem 2, the coherence of the state $|\psi_{R}\rangle$ based on the $l_{1,p}$ norm and the Tsallis relative $\alpha$ entropy, as well as that based on the $l_{1}$ norm of coherence, the skew information of coherence and the relative entropy of coherence, rely on the coefficients $\beta_{i}$ and the eigenvalues of $A$.

Applying the inverse phase estimation
$\widetilde{P}$ and measure the last qubit at the end, the state
becomes
\begin{equation*}\label{eq}
|\psi_{\widetilde{P}}\rangle=\sqrt{\frac{1}{P_{s}}}
\sum_{j=1}^{N}\beta_{j}\frac{C}{\lambda_{j}}|u_{j}\rangle.
\end{equation*}\\
$\mathbf{Theorem}$ $\mathbf{3}$ The coherence of the state $\rho_{\widetilde{P}}=|\psi_{\widetilde{P}}\rangle\langle\psi_{\widetilde{P}}|$ based on the $l_{1,p}$ norm and the Tsallis relative $\alpha$ entropy are given by
\begin{equation}\label{eq26}
C_{1,p}(\rho_{\widetilde{P}})=\sum_{i=1}^{N}\left(\sum_{j\neq i}
\left|\frac{C^{2}\beta_{i}\beta_{j}}{P_{s}\lambda_{i}\lambda_{j}}\right|^{p}\right)^{\frac{1}{p}}
\end{equation}
and
\begin{equation}\label{eq27}
C_{\alpha}(\rho_{\widetilde{P}})=\frac{1}{\alpha-1}\left[\sum_{i=1}^{N}\left(\frac{C^{2}\beta_{i}^{2}}
{P_{s}\lambda_{i}^{2}}\right)^{\frac{1}{\alpha}}-1\right],
\end{equation}
respectively,
where $C=O(\frac{1}{\kappa})$ is an appropriate constant, $P_{s}=\sum_{j=1}^{N}C^{2}\beta_{j}^{2}/\lambda_{j}^{2}$ is the success probability of HHL quantum algorithm.\\\hspace*{\fill}\\
$\it{Proof}$. Since
\begin{equation}\label{eq28}
\rho_{\widetilde{P}}=\sum_{i,j}\frac{C^{2}\beta_{i}\beta_{j}}{P_{s}\lambda_{i}\lambda_{j}}
|u_{j}\rangle\langle u_{i}|,
\end{equation}
from Eq. (\ref{eq10}) the $l_{1,p}$ norm of coherence of $\rho_{\widetilde{P}}$ is given by
\begin{equation}\label{eq29}
C_{1,p}(\rho_{\widetilde{P}})=\sum_{i=1}^{N}\left(\sum_{j\neq i}
\left|\frac{C^{2}\beta_{i}\beta_{j}}{P_{s}\lambda_{i}\lambda_{j}}\right|^{p}\right)^{\frac{1}{p}}.
\end{equation}
Since
$P_{s}=\sum_{j=1}^{N}\frac{C^{2}\beta_{j}^{2}}{\lambda_{j}^{2}}$, we
have $0<\frac{C^{2}\beta_{i}^{2}}{P_{s}\lambda_{i}^{2}}<1$. By
employing Eqs. (\ref{eq8}) and (\ref{eq28}), we obtain that the
Tsallis relative $\alpha$ entropy of coherence of the state
$|\psi_{\widetilde{P}}\rangle$ is
\begin{equation}\label{eq30}
C_{\alpha}(\rho_{\widetilde{P}})=\frac{1}{\alpha-1}\left[\sum_{i=1}^{N}\left(\frac{C^{2}\beta_{i}^{2}}
{P_{s}\lambda_{i}^{2}}\right)^{\frac{1}{\alpha}}-1\right].
\end{equation}
This completes the proof.  $\hfill\qedsymbol$ \\\hspace*{\fill}\\
{\bf Remark 3} Setting $p=1$ in Eq. (\ref{eq29}), we see that the
$l_{1}$ norm of coherence of the state
$|\psi_{\widetilde{P}}\rangle$ reduces to
\begin{equation}\label{eq31}
C_{l_{1}}(\rho_{\widetilde{P}})=\sum_{i\neq j}\frac{C^{2}|\beta_{i}\beta_{j}|}{|P_{s}\lambda_{i}\lambda_{j}|}.
\end{equation}
Letting $\alpha=\frac{1}{2}$ and taking the limit $\alpha\rightarrow1$ in Eq. (\ref{eq30}), we obtain that the skew information of coherence and the relative entropy of coherence of the state $|\psi_{\widetilde{P}}\rangle$,
\begin{equation}\label{eq32}
C_{s}(\rho_{\widetilde{P}})=1-\sum_{i=1}^{N}\left(\frac{C^{2}\beta_{i}^{2}}
{P_{s}\lambda_{i}^{2}}\right)^{2},
\end{equation}
and
\begin{equation}\label{eq33}
C_{r}(\rho_{\widetilde{P}})=-\sum_{i=1}^{N}\frac{C^{2}\beta_{i}^{2}}{P_{s}\lambda_{i}^{2}}
\log\frac{C^{2}\beta_{i}^{2}}{P_{s}\lambda_{i}^{2}},
\end{equation}
respectively,
where $\frac{C^{2}\beta_{i}^{2}}{P_{s}\lambda_{i}^{2}}\in$ (0,1).

Note that the success probablility $P_{s}$ depends on the coefficients $\beta_{i}$ and the eigenvalues of $A$. From Theorem 3 it is clear that the coherence of the state $|\psi_{\widetilde{P}}\rangle$ based on Tsallis relative $\alpha$ entropy and the $l_{1,p}$ norm rely on $\beta_{i}$ and the eigenvalues of $A$. According to Eqs. (\ref{eq31}), (\ref{eq32}) and (\ref{eq33}), the $l_{1}$ norm of coherence, the skew information of coherence and the relative entropy of coherence of the state $|\psi_{\widetilde{P}}\rangle$ also rely on these two factors.

Obviously, the $l_{1,p}$ norm of coherence and the Tsallis relative $\alpha$ entropy of coherence of the state $|\psi_{\widetilde{P}}\rangle$ both decrease with the increase of the success probability $P_{s}$ when $\alpha\in(1,2]$. However, when $\alpha\in(0,1)$, the Tsallis relative $\alpha$ entropy of coherence increases with the increase of the success probability. Setting $p=1$ and $\alpha=\frac{1}{2}$, the $l_{1}$ norm of coherence depletes with the increase of the success probability, and the skew information of coherence produces with the increase of success probability.
Note that when the success probability is so large that $C^{2}\beta_{i}^{2}/P_{s}\lambda_{i}^{2}<1/e$, the relative entropy of coherence decreases with the increase of success probability.\\\hspace*{\fill}\\
$\mathbf{Definition}$ The variation of operator coherence in one
application of HHL quantum algorithm is defined as
\begin{equation}\label{eq34}
\Delta C(\rho)\equiv C(\rho_{\widetilde{P}})-C(\rho_{P}).
\end{equation}

In HHL quantum algorithm, the quantum coherence is producing when $\Delta C(\rho)>0$, when and is depleting when $\Delta C(\rho)<0$.\\\hspace*{\fill}\\
Substituting Eqs. (\ref{eq11}), (\ref{eq12}) and (\ref{eq26}), (\ref{eq27}) into (\ref{eq34}), we have\\
$\mathbf{Theorem}$ $\mathbf{4}$ The variations of the Tsallis relative
$\alpha$ entropy of coherence and the $l_{1,p}$ norm of coherence during the
HHL quantum algorithm application are given by
\begin{equation}\label{eq35}
\Delta C_{1,p}(\rho)=\sum_{i=1}^{N}\left[\left(\sum_{j\neq i}
\left|\frac{C^{2}\beta_{i}\beta_{j}}{P_{s}\lambda_{i}\lambda_{j}}\right|^{p}
\right)^{\frac{1}{p}}-\left(\sum_{j\neq i}|\beta_{i}\beta_{j}|^{p}\right)^{\frac{1}{p}}\right],
\end{equation}
and
\begin{equation}\label{eq36}
\Delta C_{\alpha}(\rho)=\frac{1}{\alpha-1}\left[\sum_{i=1}^{N}\left(\left(\frac{C^{2}}
{P_{s}\lambda_{i}^{2}}\right)^{\frac{1}{\alpha}}-1\right)\beta_{i}^{\frac{2}{\alpha}}\right],
\end{equation}
respectively.\\\hspace*{\fill}\\
{\bf Remark 4} Letting $p=1$ in Eq. (\ref{eq35}) and
$\alpha=\frac{1}{2}$ in Eq. (\ref{eq36}), the variations of the $l_{1}$
norm of coherence and the skew information of coherence during the HHL
quantum algorithm application become
\begin{equation}\label{eq38}
\Delta C_{l_{1}}(\rho)
=\sum_{i\neq j}|\beta_{i}\beta_{j}|\left(\frac{C^{2}}{|P_{s}\lambda_{i}\lambda_{j}|}-1\right),
\end{equation}
and
\begin{equation}\label{eq39}
\Delta C_{s}(\rho)
=\sum_{i=1}^{N}\beta_{i}^{4}\left(1-\frac{C^{4}}{P_{s}^{2}\lambda_{i}^{4}}\right),
\end{equation}
respectively.
Substituting Eqs. (\ref{eq18}) and (\ref{eq33}) into (\ref{eq34}), we have the variation of the relative entropy of coherence,
\begin{equation}\label{eq40}
\Delta C_{r}(\rho)
=\sum_{i=1}^{N}\beta_{i}^{2}\log\beta_{i}^{2}-\frac{C^{2}\beta_{i}^{2}}{P_{s}\lambda_{i}^{2}}
\log\frac{C^{2}\beta_{i}^{2}}{P_{s}\lambda_{i}^{2}}.
\end{equation}

The variations of coherence deplete with the increase of success
probability based on the $l_{1,p}$ norm and
the Tsallis relative $\alpha$ entropy when
$\alpha\in(1,2]$. It is worthwhile to note that the variation of
the Tsallis relative $\alpha$ entropy of coherence
produces with the increase of the success probability when
$\alpha\in(0,1)$. Thus, the variation of the
$l_{1}$ norm of coherence depletes with the increase of success
probability, and the variation of the skew
information of coherence produces with the increase of success
probability. The variation of the relative entropy
of coherence depletes with the increase of success probability when
the success probability is so large that
$C^{2}\beta_{i}^{2}/P_{s}\lambda_{i}^{2}<1/e$.

Consider the $2\times2$ linear system of equations
$A\overrightarrow{x}=\overrightarrow{b}$ in which $A$ is a
$2\times2$ real matrix and $\overrightarrow{b}$ is a unit vector
specified by
$$A=\left(\begin{array}{cc}
a&c\\
c&d\\
\end{array}
\right),~~ \overrightarrow{b}=\left(\begin{array}{cc}
b_{0}\\
b_{1}\\
\end{array}
\right),
$$
where $c\neq0$. The following result characterizes the relationship between the coefficients $\beta_{i}$ and the elements of vector $\overrightarrow{b}$.\\\hspace*{\fill}\\
$\mathbf{Theorem}$ $\mathbf{5}$ The relationships between
coefficients $\beta_{i}$ and the elements of vector
$\overrightarrow{b}$ are given by
\begin{equation}\label{eq41}
\beta_{1}=\frac{b_{0}x_{4}-b_{1}x_{3}}{x_{1}x_{4}-x_{2}x_{3}},~~~
\beta_{2}=\frac{-b_{0}x_{2}+b_{1}x_{1}}{x_{1}x_{4}-x_{2}x_{3}}
\end{equation}
for a $2\times2$ real matrix $A$, where $x_{1}$ and $x_{2}$ ($x_{3}$ and $x_{4}$) are the coefficients obtained by decomposing the eigenvector $|u_{1}\rangle$ ($|u_{2}\rangle$) in the basis $|0\rangle$ and $|1\rangle$.\\\hspace*{\fill}\\
$\it{Proof}$. By direct calculation the eigenvalues of $A$ are
\begin{equation}\label{eq42}
\lambda_{1}=\frac{a+d-\sqrt{(a-d)^{2}+4c^{2}}}{2},~~~
\lambda_{2}=\frac{a+d+\sqrt{(a-d)^{2}+4c^{2}}}{2},
\end{equation}
with the corresponding eigenvectors
\begin{equation}\label{eq43}
|u_{1}\rangle=\frac{1}{\sqrt{1+\frac{(\lambda_{1}-a)^{2}}{c^{2}}}}|0\rangle+
\frac{\lambda_{1}-a}{c\sqrt{1+\frac{(\lambda_{1}-a)^{2}}{c^{2}}}}|1\rangle
\end{equation}
and
\begin{equation}\label{eq44}
|u_{2}\rangle=\frac{1}{\sqrt{1+\frac{(\lambda_{2}-a)^{2}}{c^{2}}}}|0\rangle+
\frac{\lambda_{2}-a}{c\sqrt{1+\frac{(\lambda_{2}-a)^{2}}{c^{2}}}}|1\rangle,
\end{equation}
respectively. Denote
$x_{1}=\frac{1}{\sqrt{1+\frac{(\lambda_{1}-a)^{2}}{c^{2}}}}$,
$x_{2}=\frac{\lambda_{1}-a}{c\sqrt{1+\frac{(\lambda_{1}-a)^{2}}{c^{2}}}}$,
$x_{3}=\frac{1}{\sqrt{1+\frac{(\lambda_{2}-a)^{2}}{c^{2}}}}$ and
$x_{4}=\frac{\lambda_{2}-a}{c\sqrt{1+\frac{(\lambda_{2}-a)^{2}}{c^{2}}}}$.
We then have
\begin{equation*}\label{eq}
|u_{1}\rangle=x_{1}|0\rangle+x_{2}|1\rangle, ~~~|u_{2}\rangle=x_{3}|0\rangle+x_{4}|1\rangle.
\end{equation*}
Since
$|b\rangle=b_{0}|0\rangle+b_{1}|1\rangle=\beta_{1}|u_{1}\rangle+\beta_{2}|u_{2}\rangle$,
we obtain that
\begin{equation*}\label{eq}
b_{0}=\beta_{1}x_{1}+\beta_{2}x_{3},~~~b_{1}=\beta_{1}x_{2}+\beta_{2}x_{4},
\end{equation*}
where $b_{0}^{2}+b_{1}^{2}=1$,
$x_{1}^{2}+x_{2}^{2}=x_{3}^{2}+x_{4}^{2}=1$,
$x_{1}x_{3}+x_{2}x_{4}=0$. Hence, it follows that
$\beta_{1}^{2}+\beta_{2}^{2}=1$ and
\begin{equation*}\label{eq}
\beta_{1}=\frac{b_{0}x_{4}-b_{1}x_{3}}{x_{1}x_{4}-x_{2}x_{3}},~~~
\beta_{2}=\frac{-b_{0}x_{2}+b_{1}x_{1}}{x_{1}x_{4}-x_{2}x_{3}}.
\end{equation*}
This completes the proof. $\hfill\qedsymbol$

In particular, for a $2\times2$ real matrix $A$, by Eq. (\ref{eq35})
the variation of the $l_{1,p}$ norm of coherence
has the following relations:
\begin{equation}\label{eq45}
\left\{
\begin{aligned}
\frac{C^{2}}{|P_{s}\lambda_{i}\lambda_{j}|}>1&;&
\left(\sum_{j\neq i}
\left|\frac{C^{2}\beta_{i}\beta_{j}}{P_{s}\lambda_{i}\lambda_{j}}\right|^{p}
\right)^{\frac{1}{p}}-\left(\sum_{j\neq i}|\beta_{i}\beta_{j}|^{p}\right)^{\frac{1}{p}}>0&,&\Delta C_{1,p}(\rho)>0,\\
\frac{C^{2}}{|P_{s}\lambda_{i}\lambda_{j}|}<1&;&
\left(\sum_{j\neq i}
\left|\frac{C^{2}\beta_{i}\beta_{j}}{P_{s}\lambda_{i}\lambda_{j}}\right|^{p}
\right)^{\frac{1}{p}}-\left(\sum_{j\neq i}|\beta_{i}\beta_{j}|^{p}\right)^{\frac{1}{p}}<0&,&\Delta C_{1,p}(\rho)<0,
\end{aligned}
\right.
\end{equation}
where $j\neq i$. Moreover, for the $l_{1}$ norm of coherence, we have
\begin{equation}\label{eq46}
\left\{
\begin{aligned}
\frac{C^{2}}{|P_{s}\lambda_{i}\lambda_{j}|}>1;~\sum_{j\neq
i}\left(\frac{C^{2}}{|P_{s}\lambda_{i}\lambda_{j}|}-1\right)|\beta_{i}\beta_{j}|>0,~~~\Delta C_{l_{1}}(\rho)>0,\\
\frac{C^{2}}{|P_{s}\lambda_{i}\lambda_{j}|}<1;~\sum_{j\neq
i}\left(\frac{C^{2}}{|P_{s}\lambda_{i}\lambda_{j}|}-1\right)|\beta_{i}\beta_{j}|<0,~~ \Delta C_{l_{1}}(\rho)<0.
\end{aligned}
\right.
\end{equation}

\vskip0.1in

\noindent {\bf 4 Examples }\\\hspace*{\fill}\\
In this section, we provide some detailed examples of linear systems
of equations for $2\times 2$ and $4\times 4$ matrices, focusing on
how coherence changes during the process of HHL quantum algorithm,
by calculating the coherence of the states and specifying the
variation of the coherence based on the Tsallis relative $\alpha$
entropy and the $l_{1,p}$ norm. \vskip0.1in \noindent {\bf Example
1} Taking the time parameter $t_{0}$ to be $2\pi$ in the
simulations, and letting $r=2$, we have $C=0.736$ which is derived
in \cite{CYD}. Since the rotation gates $R_{y}(\theta_{j})$,
$\theta_{j}=(2\pi/2^{r})\lambda_{j}^{-1}$, are a crucial part of HHL
algorithm, $r$ is an important parameter that influences the
accuracy of the solution of the linear system of equations
$A\overrightarrow{x}=\overrightarrow{b}$. For clarity, we provide a
$2\times2$ linear system of equations with
$$A=\frac{1}{2}\left(\begin{array}{cc}
3&1\\
1&3\\
\end{array}
\right),~~ \overrightarrow{b}=\frac{1}{2}\left(\begin{array}{cc}
\sqrt{3}\\
1\\
\end{array}
\right).
$$
$A$ is hermitian with eigenvalues $\lambda_{1}=1$ and
$\lambda_{2}=2$ and the corresponding eigenvectors
$\overrightarrow{u_{1}}=\frac{1}{\sqrt{2}}\left(\begin{matrix}1\\-1\end{matrix}\right)$
and
$\overrightarrow{u_{2}}=\frac{1}{\sqrt{2}}\left(\begin{matrix}1\\1\end{matrix}
\right)$. Then
$|u_{1}\rangle=\frac{1}{\sqrt{2}}|0\rangle-\frac{1}{\sqrt{2}}|1\rangle$
and
$|u_{2}\rangle=\frac{1}{\sqrt{2}}|0\rangle+\frac{1}{\sqrt{2}}|1\rangle$,
and we have $x_{1}=x_{3}=x_{4}=\frac{1}{\sqrt{2}}$ and
$x_{2}=-\frac{1}{\sqrt{2}}$. Expressing $\overrightarrow{b}$ as a
quantum state
$|b\rangle=\frac{\sqrt{3}}{2}|0\rangle+\frac{1}{2}|1\rangle$ with
$b_{0}=\frac{\sqrt{3}}{2}$ and $b_{1}=\frac{1}{2}$, and applying the
phase estimation and decomposing $|b\rangle$ in the eigenvector
basis, according to Eq. (\ref{eq41}) we have
$\beta_{1}=\frac{\sqrt{6}-\sqrt{2}}{4}$ and
$\beta_{2}=\frac{\sqrt{6}+\sqrt{2}}{4}$. Thus the quantum state
$|b\rangle$ can be reexpressed as
\begin{align}\label{eq47}
|b\rangle=\sum_{j=1}^{2}\beta_{j} |u_{j}\rangle=\frac{\sqrt{6}-\sqrt{2}}{4}|u_{1}\rangle+\frac{\sqrt{6}+\sqrt{2}}{4}|u_{2}\rangle.
\end{align}
We get the value of success probability
$P_{s}=\sum_{j=1}^{N}\frac{C^{2}\beta_{j}^{2}}{\lambda_{j}^{2}}\approx0.162$,
and the value
$\frac{C^{2}}{P_{s}\lambda_{i}\lambda_{j}}\approx1.672>1$. According
to Eq. (\ref{eq45}) we have $\Delta C_{1,p}(\rho)>0$, the variation of the
$l_{1,p}$ norm
of coherence produces in one application of HHL quantum algorithm.\\
$\bullet$ (i) By Eqs. (\ref{eq11}) and (\ref{eq12}), after the phase estimation the coherence of the state $|\psi_{P}\rangle$ based on the $l_{1,p}$ norm and the Tsallis relative $\alpha$ entropy are
\begin{equation}\label{eq48}
C_{1,p}(\rho_{P})=\frac{1}{2}
\end{equation}
and
\begin{equation}\label{eq49}
C_{\alpha}(\rho_{P})\approx\frac{1}{\alpha-1}\left[(0.067)^{\frac{1}{\alpha}}+
(0.933)^{\frac{1}{\alpha}}-1\right],
\end{equation}
respectively. The $l_{1}$ norm of coherence, the
 skew information of coherence and the relative entropy of coherence of the state $|\psi_{P}\rangle$ are
\begin{equation*}\label{eq}
C_{l_{1}}(\rho_{P})=\frac{1}{2},~~
C_{s}(\rho_{P})\approx0.125~~~\mathrm{and}~~~
C_{r}(\rho_{P})\approx0.355,
\end{equation*}
respectively.\\
$\bullet$ (ii) By Eq. (\ref{eq19}), after conditional rotation $C$-$R_{y}(\theta_{j})$ the coherence of the state $|\psi_{R}\rangle$ based on the $l_{1,p}$ norm can be expressed as
\begin{align}\label{eq50}
C_{1,p}(\rho_{R})\notag
\approx&\left((0.157)^{p}+(0.033)^{p}+(0.062)^{p}\right)^{\frac{1}{p}}
+\left((0.157)^{p}+(0.171)^{p}+(0.319)^{p}\right)^{\frac{1}{p}}\\
+&\left((0.068)^{p}+(0.033)^{p}+(0.171)^{p}\right)^{\frac{1}{p}}+
\left((0.068)^{p}+(0.062)^{p}+(0.319)^{p}\right)^{\frac{1}{p}},
\end{align}
and by Eq. (\ref{eq20}), the coherence of the state $|\psi_{R}\rangle$ based on the Tsallis relative $\alpha$ entropy is given by
\begin{equation}\label{eq51}
C_{\alpha}(\rho_{R})
\approx\frac{1}{\alpha-1}\left[(0.031)^{\frac{1}{\alpha}}+(0.036)^{\frac{1}{\alpha}}
+(0.807)^{\frac{1}{\alpha}}+(0.126)^{\frac{1}{\alpha}}-1\right].
\end{equation}
Moreover, the $l_{1}$ norm of coherence, the skew information of coherence and the relative entropy of coherence of $\rho_{R}$ are
\begin{equation*}\label{eq}
C_{l_{1}}(\rho_{R})\approx1.620,~~~
C_{s}(\rho_{R})\approx0.331~~~\mathrm{and}~~~
C_{r}(\rho_{R})\approx0.954,
\end{equation*}
respectively.\\
$\bullet$ (iii) By Eqs. (\ref{eq26}) and (\ref{eq27}), after the inverse phase estimation and measurement the coherence of the state $|\psi_{\widetilde{P}}\rangle$ based on $l_{1,p}$ norm and Tsallis relative $\alpha$ entropy are given by
\begin{equation}\label{eq52}
C_{1,p}(\rho_{\widetilde{P}})\approx0.836
\end{equation}
and
\begin{equation}\label{eq53}
C_{\alpha}(\rho_{\widetilde{P}})\approx\frac{1}{\alpha-1}\left[(0.224)^{\frac{1}{\alpha}}+
(0.780)^{\frac{1}{\alpha}}-1\right],
\end{equation}
respectively. In addition, the coherence of the state $|\psi_{\widetilde{P}}\rangle$ based on the $l_{1}$ norm, the skew information and the relative entropy are
\begin{equation*}\label{eq}
C_{l_{1}}(\rho_{\widetilde{P}})\approx0.836,~~~
C_{s}(\rho_{\widetilde{P}})\approx0.341~~~\mathrm{and}~~~
C_{r}(\rho_{\widetilde{P}})\approx0.763,
\end{equation*}
respectively.\\
$\bullet$ (iv) The variations of coherence based on the $l_{1,p}$ norm
and the Tsallis relative $\alpha$ entropy during the HHL quantum
algorithm application are
\begin{equation}\label{eq54}
\Delta C_{1,p}(\rho)\approx 0.336
\end{equation}
and
\begin{equation}\label{eq55}
\Delta C_{\alpha}(\rho)=\frac{1}{\alpha-1}\left[(0.224)^{\frac{1}{\alpha}}+
(0.780)^{\frac{1}{\alpha}}-(0.067)^{\frac{1}{\alpha}}-
(0.933)^{\frac{1}{\alpha}}\right],
\end{equation}
respectively. Accordingly, the variations of coherence based on the $l_{1}$ norm, the skew information and the relative entropy are
\begin{equation*}\label{eq}
\Delta C_{l_{1}}(\rho)\approx 0.336,~~~ \Delta
C_{s}(\rho)=0.216~~~\mathrm{and}~~~ \Delta C_{r}(\rho)=0.408,
\end{equation*}
respectively.

It is clear that the coherence based on the $l_{1}$ norm and the
relative entropy first increase and then decrease, and the skew
information of coherence always increases when the phase estimation
$P$, conditional rotation $C$-$R_{y}(\theta_{j})$ and inverse phase
estimation $\widetilde{P}$ are applied in HHL quantum algorithm.

\begin{figure}[H]\centering
\subfigure[] {\begin{minipage}[figure1a]{0.49\linewidth}
\includegraphics[width=1.0\textwidth,natwidth=12cm,natheight=9cm]{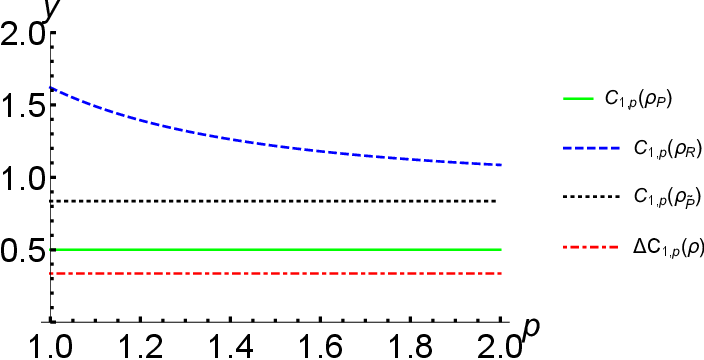}
\end{minipage}}
\subfigure[] {\begin{minipage}[figure1b]{0.49\linewidth}
\includegraphics[width=1.0\textwidth,natwidth=12cm,natheight=9cm]{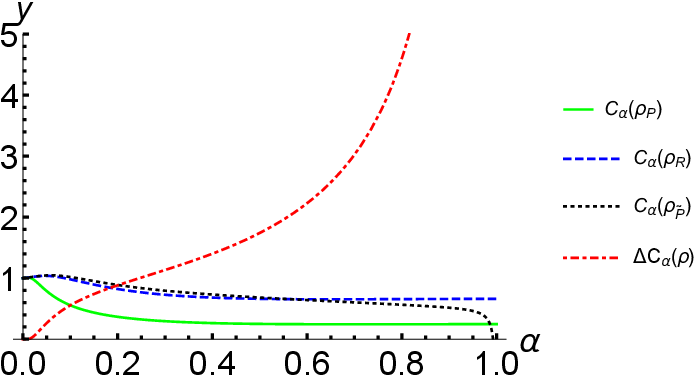}
\end{minipage}}
\caption{The $y$-axis stands for the values of coherence.
{Subfigures $\mathbf{a}$ ($\mathbf{b}$) is for the case that the
coherence based on the $l_{1,p}$ norm (Tsallis
relative $\alpha$ entropy). The operator coherence of $P$ (green,
Eqs. (\ref{eq48}) and (\ref{eq49})), $R$ (blue dashed, Eqs.
(\ref{eq50}) and (\ref{eq51})), $\widetilde{P}$ (black dotted, Eqs.
(\ref{eq52}) and (\ref{eq53})) and the variations of coherence based
on the $l_{1,p}$ norm and the
Tsallis relative $\alpha$ entropy (red dot-dashed, Eqs. (\ref{eq54})
and (\ref{eq55})). \label{fig:Fig1}}}
\end{figure}

The quantities given by Eqs. (\ref{eq48})-(\ref{eq55}) are
illustrated in Fig. 1 for $p\in [1,2]$ and $\alpha\in (0,1)$.
According to Example 1 and Fig. 1, it is observed that the $l_{1,p}$
norm of coherence of $P$ and $\widetilde{P}$, and thus the variation
of coherence in one application of the algorithm, are all
independent of the value of $p$. The operator coherence based on
the Tsallis relative $\alpha$ entropy, however, all depends on $\alpha$,
and the coherence of each operator decreases as $\alpha$ increases
when $\alpha\in[\frac{1}{10},1]$. Moreover, it holds that
$C_{1,p}(\rho_P)\leq C_{1,p}(\rho_{\widetilde{P}})\leq
C_{1,p}(\rho_R)$ for each $p$, but $C_{\alpha}(\rho_P)\leq
C_{\alpha}(\rho_{\widetilde{P}})\leq C_{\alpha}(\rho_R)$ does not hold for each $\alpha$. Nevertheless, in HHL quantum algorithm, the overall effect based on different coherence measures are both coherence production for the same $2\times2$ linear system of equations.\\\hspace*{\fill}\\
\noindent {\bf Example 2} Consider a $4\times4$ linear system of
equations $A\overrightarrow{x}=\overrightarrow{b}$ specified by
$$A=\left(\begin{array}{cc}
1&0~~0~~0\\
0&2~~0~~0\\
0&0~~3~~0\\
0&0~~0~~4\\
\end{array}
\right),~~ \overrightarrow{b}=\frac{1}{4}\left(\begin{array}{cc}
\sqrt{10}\\
\sqrt{2}\\
0\\
2\\
\end{array}
\right).
$$
The matrix $A$ is hermitian with the eigenvalues $\lambda_{1}=1$,
$\lambda_{2}=2$, $\lambda_{3}=3$ and $\lambda_{4}=4$, and the
corresponding eigenvectors
$$
\overrightarrow{u_{1}}=\left(\begin{matrix}1\\0\\0\\0\end{matrix}\right),~~~
\overrightarrow{u_{2}}=\left(\begin{matrix}0\\1\\0\\0\end{matrix}\right),~~~
\overrightarrow{u_{3}}=\left(\begin{matrix}0\\0\\1\\0\end{matrix}\right),~~~
\overrightarrow{u_{4}}=\left(\begin{matrix}0\\0\\0\\1\end{matrix}\right).
$$
Expressing $\overrightarrow{b}$ as a quantum state
$|b\rangle=\frac{\sqrt{10}}{4}|00\rangle+\frac{\sqrt{2}}{4}|01\rangle+0|10\rangle
+\frac{1}{2}|11\rangle$, we get $|u_{1}\rangle=|00\rangle$,
$|u_{2}\rangle=|01\rangle$, $|u_{3}\rangle=|10\rangle$ and
$|u_{4}\rangle=|11\rangle$. Applying the phase estimation and
decomposing $|b\rangle$ in the eigenvector basis,
\begin{align}\label{eq56}
|b\rangle=\sum_{j=1}^{4}\beta_{j} |u_{j}\rangle=\frac{\sqrt{10}}{4}|u_{1}\rangle+\frac{\sqrt{2}}{4}|u_{2}\rangle+0|u_{3}\rangle
+\frac{1}{2}|u_{4}\rangle,
\end{align}
we have $\beta_{1}=\frac{\sqrt{10}}{4}$, $\beta_{2}=\frac{\sqrt{2}}{4}$, $\beta_{3}=0$ and $\beta_{4}=\frac{1}{2}$. Taking the time parameter $t_0=2\pi$, $r=2$, we obtain $C=0.736$ derived from \cite{CYD}, and thus the success probability $P_{s}=\sum_{j=1}^{N}C^{2}\beta_{j}^{2}/\lambda_{j}^{2}\approx0.364$.\\
$\bullet$ (i) By Eqs. (\ref{eq11}) and (\ref{eq12}), after the phase
estimation the coherence of the state $|\psi_{P}\rangle$ based on the
$l_{1,p}$ norm and the Tsallis relative $\alpha$ entropy are
\begin{equation}\label{eq57}
C_{1,p}(\rho_{P})\approx\left((0.280)^{p}+(0.395)^{p}\right)^{\frac{1}{p}}
+\left((0.280)^{p}+(0.177)^{p}\right)^{\frac{1}{p}}
+\left((0.395)^{p}+(0.177)^{p}\right)^{\frac{1}{p}}
\end{equation}
and
\begin{equation}\label{eq58}
C_{\alpha}(\rho_{P})\approx\frac{1}{\alpha-1}\left[(0.625)^{\frac{1}{\alpha}}+
(0.125)^{\frac{1}{\alpha}}+(0.250)^{\frac{1}{\alpha}}-1\right],
\end{equation}
respectively. The coherence of the state $|\psi_{P}\rangle$ based on the $l_{1}$ norm, the skew information and the relative entropy are
\begin{equation*}\label{eq}
C_{l_{1}}(\rho_{P})\approx1.704,~~
C_{s}(\rho_{P})\approx0.531~~~\mathrm{and}~~~
C_{r}(\rho_{P})\approx1.299,
\end{equation*}
respectively.\\
$\bullet$ (ii) By Eq. (\ref{eq19}), after conditional rotation $C$-$R_{y}(\theta_{j})$ the coherence of the state $|\psi_{R}\rangle$ based on the $l_{1,p}$ norm can be expressed as
\begin{align}\label{eq59}
C_{1,p}(\rho_{R})\notag
\approx&\left((0.176)^{p}+(0.263)^{p}+(0.311)^{p}+(0.070)^{p}+(0.049)^{p}\right)^{\frac{1}{p}}\\ \notag
+&\left((0.176)^{p}+(0.162)^{p}+(0.191)^{p}+(0.043)^{p}+(0.030)^{p}\right)^{\frac{1}{p}}\\ \notag
+&\left((0.263)^{p}+(0.162)^{p}+(0.286)^{p}+(0.064)^{p}+(0.045)^{p}\right)^{\frac{1}{p}}\\ \notag
+&\left((0.076)^{p}+(0.053)^{p}+(0.311)^{p}+(0.191)^{p}+(0.286)^{p}\right)^{\frac{1}{p}}\\ \notag
+&\left((0.076)^{p}+(0.012)^{p}+(0.070)^{p}+(0.043)^{p}+(0.064)^{p}\right)^{\frac{1}{p}}\\
+&\left((0.053)^{p}+(0.012)^{p}+(0.049)^{p}+(0.030)^{p}+(0.045)^{p}\right)^{\frac{1}{p}},
\end{align}
and by Eq. (\ref{eq20}), we have the coherence of the state $|\psi_{R}\rangle$ based on the Tsallis relative $\alpha$ entropy,
\begin{equation}\label{eq60}
C_{\alpha}(\rho_{R})
\approx\frac{1}{\alpha-1}\left[(0.286)^{\frac{1}{\alpha}}+(0.339)^{\frac{1}{\alpha}}
+(0.108)^{\frac{1}{\alpha}}
+(0.017)^{\frac{1}{\alpha}}+(0.242)^{\frac{1}{\alpha}}+(0.009)^{\frac{1}{\alpha}}-1\right].
\end{equation}
The $l_{1}$ norm of coherence, the skew information of coherence and the relative entropy of coherence of $\rho_{R}$ are
\begin{equation*}\label{eq}
C_{l_{1}}(\rho_{R})\approx3.662,~~~
C_{s}(\rho_{R})\approx0.733~~~\mathrm{and}~~~
C_{r}(\rho_{R})\approx2.049.
\end{equation*}
respectively.\\
$\bullet$ (iii) By Eqs. (\ref{eq26}) and (\ref{eq27}), after the inverse phase estimation and measurement the coherence of the state $|\psi_{\widetilde{P}}\rangle$ based on the $l_{1,p}$ norm and the Tsallis relative $\alpha$ entropy can be expressed as
\begin{equation}\label{eq61}
C_{1,p}(\rho_{\widetilde{P}})\approx\left((0.208)^{p}+(0.147)^{p}\right)^{\frac{1}{p}}
+\left((0.208)^{p}+(0.033)^{p}\right)^{\frac{1}{p}}
+\left((0.147)^{p}+(0.033)^{p}\right)^{\frac{1}{p}}
\end{equation}
and
\begin{equation}\label{eq62}
C_{\alpha}(\rho_{\widetilde{P}})\approx\frac{1}{\alpha-1}\left[(0.930)^{\frac{1}{\alpha}}+
(0.047)^{\frac{1}{\alpha}}+(0.023)^{\frac{1}{\alpha}}-1\right],
\end{equation}
respectively. The coherence of the state $|\psi_{\widetilde{P}}\rangle$ based on the $l_{1}$ norm, the skew information and the relative entropy are given by
\begin{equation*}\label{eq}
C_{l_{1}}(\rho_{\widetilde{P}})\approx0.776,~~~
C_{s}(\rho_{\widetilde{P}})\approx0.132~~~\mathrm{and}~~~
C_{r}(\rho_{\widetilde{P}})\approx0.430,
\end{equation*}
respectively.\\
$\bullet$ (iv) The variations of coherence based on the $l_{1,p}$ norm
and the Tsallis relative $\alpha$ entropy during the HHL quantum
algorithm application are
\begin{align}\label{eq63}
\Delta C_{1,p}(\rho)\notag
=&\left((0.208)^{p}+(0.147)^{p}\right)^{\frac{1}{p}}
+\left((0.208)^{p}+(0.033)^{p}\right)^{\frac{1}{p}}+\left((0.147)^{p}
+(0.033)^{p}\right)^{\frac{1}{p}}\\
-&\left((0.280)^{p}+(0.395)^{p}\right)^{\frac{1}{p}}
-\left((0.280)^{p}+(0.177)^{p}\right)^{\frac{1}{p}}-\left((0.395)^{p}
+(0.177)^{p}\right)^{\frac{1}{p}}
\end{align}
and
\begin{equation}\label{eq64}
\Delta C_{\alpha}(\rho)=\frac{1}{\alpha-1}\left[(0.930)^{\frac{1}{\alpha}}+
(0.047)^{\frac{1}{\alpha}}+(0.023)^{\frac{1}{\alpha}}-(0.625)^{\frac{1}{\alpha}}-
(0.125)^{\frac{1}{\alpha}}-(0.250)^{\frac{1}{\alpha}}\right],
\end{equation}
respectively. The variations of coherence based on $l_{1}$ norm, skew information and relative entropy are
\begin{equation*}\label{eq}
\Delta C_{l_{1}}(\rho)=-0.928,~~~
\Delta C_{s}(\rho)=-0.399~~~\mathrm{and}~~~
\Delta C_{r}(\rho)=-0.869,
\end{equation*}
respectively.

It is easy to see that the coherence based on the
$l_{1}$ norm, the skew information and the relative entropy first increases
and then decreases when $P$, $R$ and $\widetilde{P}$ are applied.
The quantities in (\ref{eq57})-(\ref{eq64}) are depicted in Fig. 2
for $p\in [1,2]$ and $\alpha\in (0,1)$, and it is shown that the
$l_{1,p}$ norm of coherence of each operator decreases as $p$
increases, and the magnitude of the decrease are slightly different.
The Tsallis relative $\alpha$ entropy of coherence depends on
$\alpha$ and the trend of change varies for distinct operators.
Moreover, unlike Example 1, it can be found that
$C_{1,p}(\rho_{\widetilde{P}})\leq C_{1,p}(\rho_P) \leq
C_{1,p}(\rho_R)$ for each $p$, and
$C_{\alpha}(\rho_{\widetilde{P}})\leq C_{\alpha}(\rho_P) \leq
C_{\alpha}(\rho_R)$ for each $\alpha$, i.e., the same orderings hold
no matter which coherence quantifier is used. Notably, the overall
effect in this example based on two different coherence measures are
both coherence depletion for the same linear system of equations.

\begin{figure}[H]\centering
\subfigure[] {\begin{minipage}[figure2a]{0.49\linewidth}
\includegraphics[width=1.0\textwidth,natwidth=12cm,natheight=9cm]{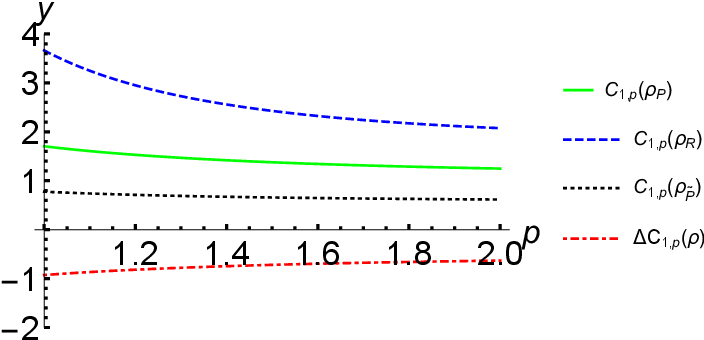}
\end{minipage}}
\subfigure[] {\begin{minipage}[figure2b]{0.49\linewidth}
\includegraphics[width=1.0\textwidth,natwidth=12cm,natheight=9cm]{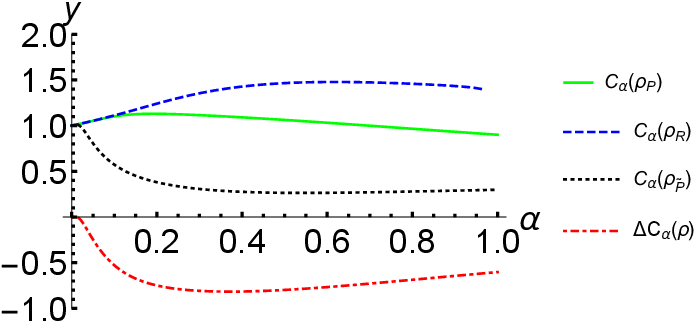}
\end{minipage}}
\caption{The $y$-axis stands for the values of coherence.
{Subfigures $\mathbf{a}$ ($\mathbf{b}$) is for the case that the
coherence based on the $l_{1,p}$ norm (Tsallis
relative $\alpha$ entropy). The operator coherence of $P$ (green,
Eqs. (\ref{eq57}) and (\ref{eq58})), $R$ (blue dashed, Eqs.
(\ref{eq59}) and (\ref{eq60})), $\widetilde{P}$ (black dotted, Eqs.
(\ref{eq61}) and (\ref{eq62})) and the variations of coherence based
on the $l_{1,p}$ norm and the
Tsallis relative $\alpha$ entropy (red dot-dashed, Eqs. (\ref{eq63})
and (\ref{eq64})). \label{fig:Fig2}}}
\end{figure}


The quantum mechanical systems differ significantly from classical
ones mainly due to coherence, so it is useful to see how coherence
changes with the execution of a program for quantum algorithms. From
Examples 1 and 2, it can be found that the overall results of
different coherence measures rely on the selection of matrix $A$ and
vector $\overrightarrow{b}$, and for the same linear system of
equations, the coherence in the whole process of HHL quantum
algorithm is producing or depleting regardless of the chosen
coherence quantifier.

Based on the theoretical framework of HHL quantum algorithm
outlined, the phase estimation ideally gives the state
$\sum_{j=1}^{N}\beta_{j}|\lambda_{j}\rangle|u_{j}\rangle$ in
register $D$ and $B$. After conditional rotation and inverse phase
estimation, the success probability is
$\sum_{j=1}^{N}C^{2}\beta_{j}^{2}/\lambda_{j}^{2}$. Given a
particular system of linear equations
$A\overrightarrow{x}=\overrightarrow{b}$, the coefficients
$\beta_{i}$, eigenvalues and success probability can be determined.
For an arbitrary $2\times2$ real matrix $A$, it can be deduced that
the coherence based on the $l_{1,p}$ norm of $P$ and $\widetilde{P}$
are always $2|\beta_{i}\beta_{j}|$ and
$2|\frac{C^{2}\beta_{i}\beta_{j}}{P_{s}\lambda_{i}\lambda_{j}}|$,
respectively. Therefore the coherence of $P$ and $\widetilde{P}$, as
well as the variations of coherence, are always constants and have
nothing to do with $p$. When the dimension is greater than 2, this
rule no longer holds, and the amount of the $l_{1,p}$ norm of
coherence is related to the value of $p$. For a particular system of
linear equations, the Tsallis relative $\alpha$ entropy of coherence
of $P$, $R$ and $\widetilde{P}$ only depends on $\alpha$,
irrespective of the form of the matrix $A$. Undoubtedly the
coherence dynamics in the HHL algorithm are determined by both the
coherence quantifier and the system of linear equation taken into
account.

\vskip0.1in

\noindent {\bf 5 Conclusions and discussions}\\\hspace*{\fill}\\
We have studied how each basic operator contributes to the Tsallis
relative $\alpha$ entropy of coherence and the $l_{1,p}$ norm of
coherence in HHL quantum algorithm, proving that the coherence of
the phase estimation $P$ relies on the coefficients $\beta_{i}$
obtained by decomposing $|b\rangle$ in the eigenbasis of $A$.
Moreover, we have displayed that the operator coherence of
controlled rotation $R_{y}(\theta_{j})$ and inverse phase estimation
$\widetilde{P}$ both rely on the coefficients $\beta_{i}$ and the
eigenvalues of $A$. The Tsallis relative $\alpha$ entropy of
coherence and the $l_{1,p}$ norm of coherence of inverse phase
estimation $\widetilde{P}$ decrease with the increase of the success
probability when $\alpha\in(1,2]$, while the Tsallis relative
$\alpha$ entropy of coherence increases with the increase of success
probability when $\alpha\in(0,1)$.

In addition, we have studied the coherence production and depletion
in HHL quantum algorithm, showing that the variations of coherence
deplete with the increase of the success probability based on the
$l_{1,p}$ norm and the Tsallis relative $\alpha$ entropy when
$\alpha\in(1,2]$. Moreover, we have proved that the produced or
depleted coherence relies on the eigenvalues of $A$ and the success
probability, and the relationships between the coefficients
$\beta_{i}$ and the elements of vector $\overrightarrow{b}$ have
been analytically derived for any $2\times 2$ real matrix $A$.

It is argued in \cite{LYC} that the coherence in Deutsch-Jozsa
algorithm, Shor's algorithm and Grover's algorithm are always
consumed, but similar assertions do not hold in these algorithms for
other quantifiers of quantum resources, such as quantum
entanglement, possibly because the definitions of coherence is
basis-dependent, while the one of entanglement is not. As a result,
it is believed that \cite{LYC} coherence depletion is a common
feature in these algorithms, which might be beneficial to design new
quantum algorithms. Surprisingly, the novel perspective we get here
is that quantum coherence in HHL quantum algorithm is not only
consumed, but also produced. We may apply the same method to more
quantum algorithms to study the coherence dynamics, and the
relationship between coherence and success probability of the
algorithm. In this regard, our results may shed some new light on
the study of coherence dynamics in other quantum algorithms, and
provide new insights into quantum computing.

\vskip0.1in

\noindent

\subsubsection*{Acknowledgements}
\small {This work was supported by National Natural Science
Foundation of China (Grant Nos. 12161056, 12075159, 12171044);
Natural Science Foundation of Jiangxi Province (Grant No.
20232ACB211003); Beijing Natural Science Foundation (Grant No.
Z190005); the Academician Innovation Platform of Hainan Province.}


\subsubsection*{Data availability statement}
\small {No new data were created or analysed in this study.}


\subsubsection*{Conflict of interest}
\small {On behalf of all authors, the corresponding author states
that there is no conflict of interest.}


\end{document}